\documentclass[10pt,journal,compsoc]{IEEEtran}

\usepackage[nocompress]{cite}

\usepackage[pdftex]{graphicx}
\usepackage{enumitem}
\setlist[enumerate]{itemsep=0mm}

\usepackage{color}
\usepackage{subfigure}
\graphicspath{{figure/}{pictures/}{images/}{./}} 
\usepackage{hyperref}
\usepackage{cite}
\usepackage{graphicx}
\usepackage{amsmath}

\usepackage{booktabs} 
\usepackage{multirow}
\usepackage{array}
\usepackage{caption}

\usepackage[table]{xcolor}
\usepackage{cite}
\usepackage{amsmath,amssymb,amsfonts}
\usepackage{textcomp}
\usepackage{xcolor}

\usepackage{color}
\usepackage{subfigure}
\usepackage[ruled]{algorithm2e}

\usepackage[T1]{fontenc}
\usepackage{fix-cm}

\usepackage{subfigure}
\usepackage{caption}

\begin{document}

\title{Discovering Opioid Use Patterns from Social Media for Relapse Prevention}

\author{
        Zhou~Yang,
        Spencer Bradshaw,
        Rattikorn Hewett,
        and~Fang~Jin
\IEEEcompsocitemizethanks{
\IEEEcompsocthanksitem Zhou Yang, Rattikorn Hewett, and Fang Jin are with the Department of Computer Science, Texas Tech University, Lubbock, Texas, USA.
E-mail: \text{\{zhou.yang, rattikorn.hewett, fang.jin\}@ttu.edu}
\IEEEcompsocthanksitem Spencer Bradshaw is with the Community, Family, and Addiction Sciences, Texas Tech University, Lubbock, Texas, USA. Email: spencer.bradshaw@ttu.edu
}
}

\IEEEtitleabstractindextext{%
    \begin{abstract}
    The United States is currently experiencing an unprecedented opioid crisis, and opioid overdose has become a leading cause of injury and death. Effective opioid addiction recovery calls for not only medical treatments, but also behavioral interventions for impacted individuals. In this paper, we study communication and behavior patterns of patients with opioid use disorder (OUD) from social media, intending to demonstrate how existing information from common activities, such as online social networking, might lead to better prediction, evaluation, and ultimately prevention of relapses. Through a multi-disciplinary and advanced novel analytic perspective, we characterize opioid addiction behavior patterns by analyzing opioid groups from Reddit.com - including modeling online discussion topics, analyzing text co-occurrence and correlations, and identifying emotional states of people with OUD. These quantitative analyses are of practical importance and demonstrate innovative ways to use information from online social media, to create technology that can assist in relapse prevention.
    \end{abstract}

}

\maketitle

\IEEEdisplaynontitleabstractindextext

\IEEEpeerreviewmaketitle


\section{Introduction}
\begin{figure*}[t]
	\centering
	\subfigure[]{
	   \includegraphics[width=0.45\linewidth]{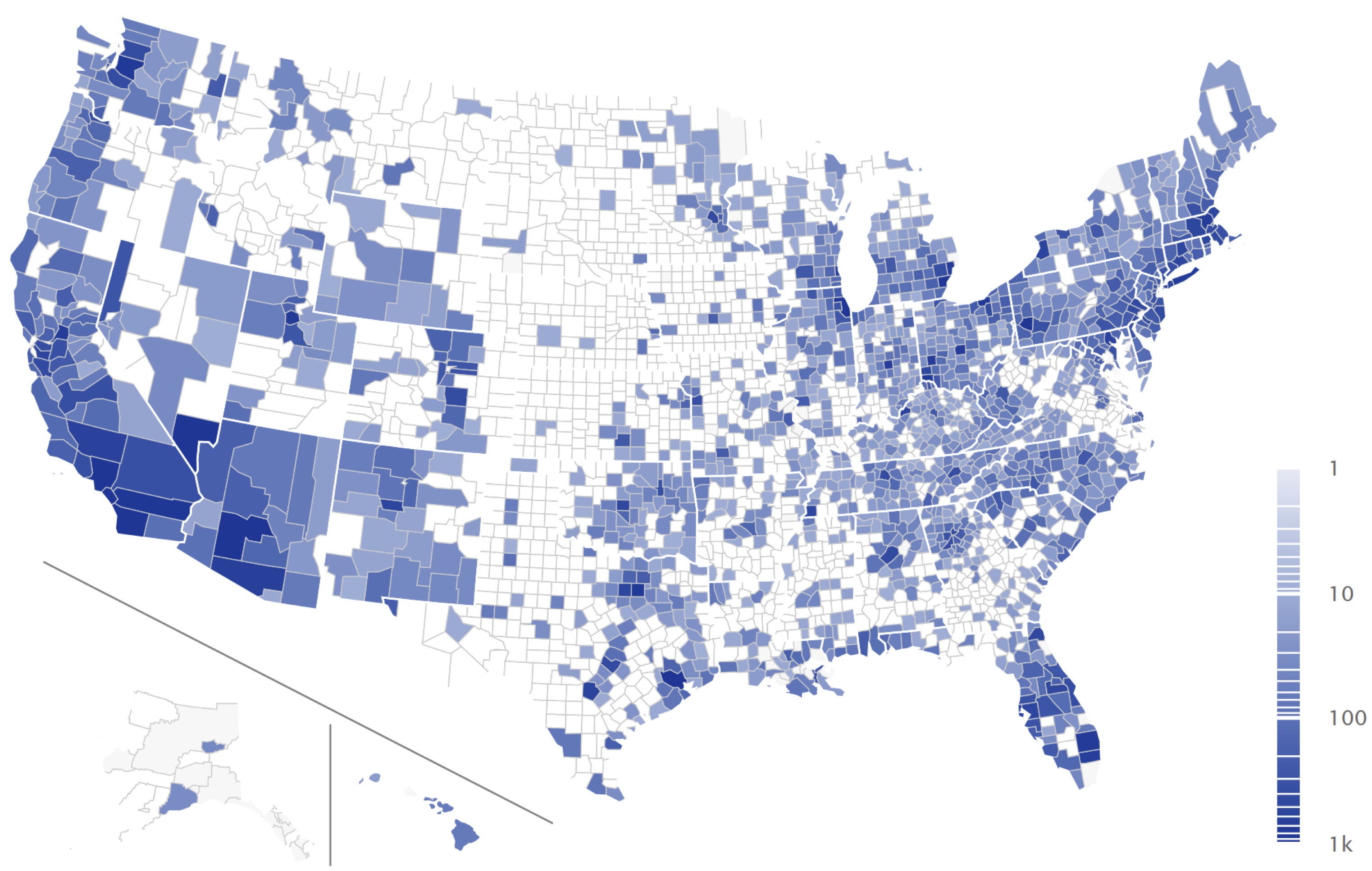}
	}
	\subfigure[]{
	   \includegraphics[width=0.45\linewidth]{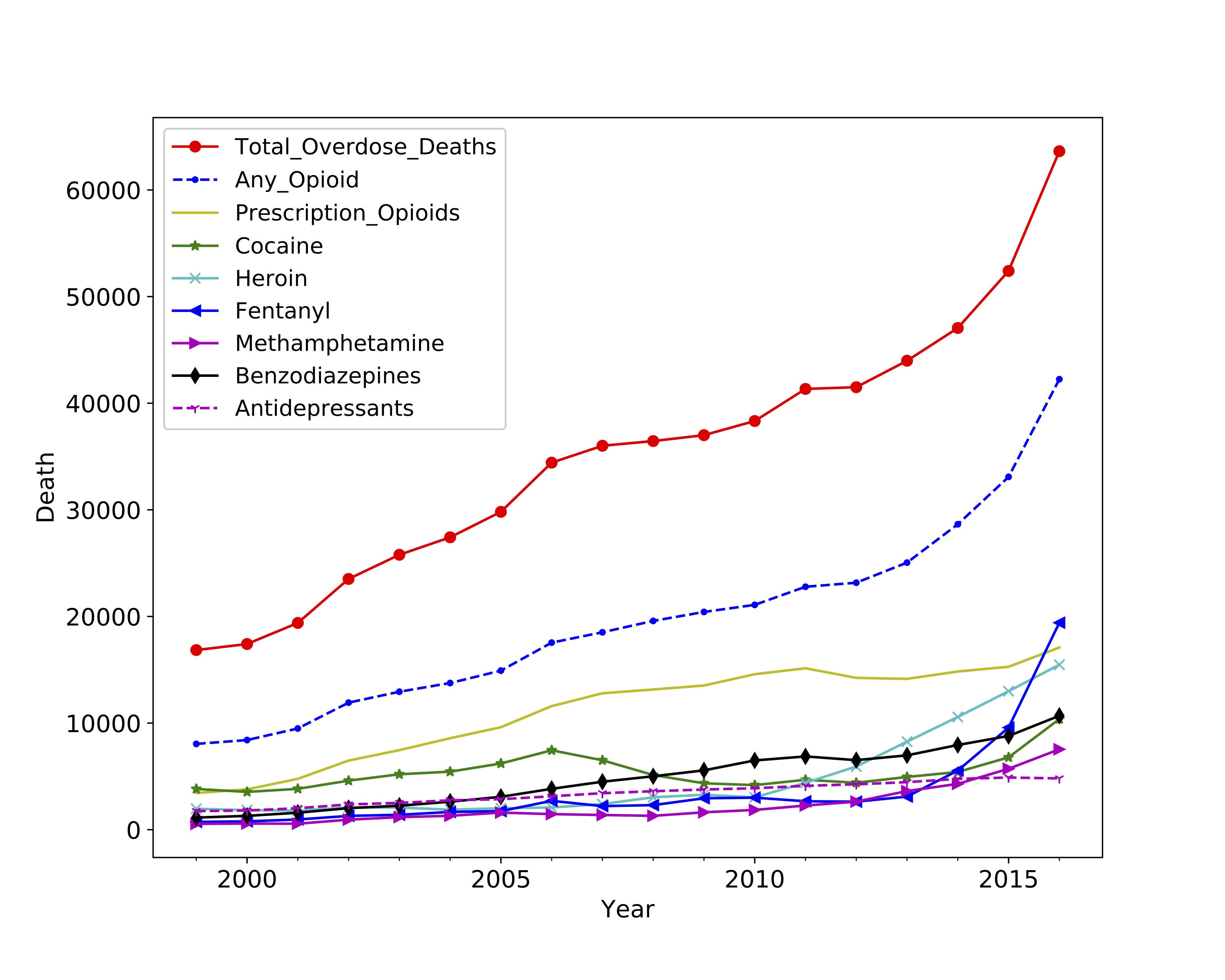}
	}
	
	\caption{\textbf{(a).Drug overdose death toll by county in 2017~\cite{Death_rates}; }\textbf{(b).Drug overdose death toll by drug type from 1996 to 2017 in the U.S.}}
	\label{fig:map_stats}
\end{figure*}

Opioid overdoses are now causing more deaths than car crashes, prompting the current U.S. President to declare the opioid crisis a national public health emergency in October 2017. According to the latest statistics from the National Institute on Drug Abuse (NIDA), more than 115 Americans die daily after overdosing on opioids, and nearly 64,000 people died of drug overdoses in the US in 2016, the most lethal year of the drug overdose epidemic~\cite{cdc_drugoverdose}. Moreover, millions of Americans have been impacted by opioid-related problems. It is estimated that 2.1 million people suffer from substance use disorders related to prescription opioid pain relievers in the United States alone~\cite{drugabusegov}. Additionally, the opioid crisis has social impacts beyond an increased death toll. Other consequences include a rise in the number of infants born dependent on opioids as well as a spread of infectious diseases such as HIV and hepatitis C~\cite{yang2019asonam}. The status of the opioid crisis in the U.S. is shown in Figure~\ref{fig:map_stats} (a) and Figure~\ref{fig:map_stats} (b) from different perspectives. As revealed in a recent paper published in Science~\cite{jalal2018changing}, the drug epidemic currently sweeping across the U.S. has deteriorated into a full-scale national pandemic, leading to national concern because of its negative impacts on health, social security and economics.

Current knowledge of the opioid crisis has been mostly limited to scattered statistics that only reveal macro aspects of the crisis, such as the nationwide/state-level overdose death tolls and overdose death tolls regarding certain time periods, races, and/or specific drugs. This has led to a rapid increase in OUD treatment admissions, including a $53.5\%$ increase during 2013 - 2015, and with adult treatment for heroin use more than doubling~\cite{Huhn2018TreatmentAdmission}. While treatment is important and leads to effective outcomes and greater health for some, the relapse rate post-treatment for substance use disorders is relatively high, ranging from $40-60\%$~\cite{McLellan2000ChronicIllness}, with potential relapse being difficult to predict. Therefore, research that can enhance relapse prediction using data from everyday, common situations is needed.

One such resource may include behavioral data available from social media, a forum now used by $88\%$ of persons in the U.S. ages 18-29, $78\%$ of those ages 30-49, $64\%$ of those ages 50-64, and $37\%$ of those age 65 and older~\cite{Smith2018SocialMediaUse}. Despite its popularity, detailed analysis of user-specific knowledge, such as social media behavioral patterns that associate with substance use and relapse, have not been studied. Therefore, to facilitate a better understanding of the opioid epidemic that may ultimately lead to better prediction of relapse and ultimately responses and interventions that prevent relapse and promote remission - individualized, user-specific analysis is urgently needed. This manuscript begins to tackle this complex problem from a multi-disciplinary perspective, and includes text analysis, topic modeling,and sentiment analysis.

\section{Data Analysis and Results}
\subsection{Opioid Communities on Social Media}
With social media growing profoundly and widely intertwined with people's everyday lives, social media contains more and more biomedical psychological information about its users. As a result, social media represents an increasingly important data source for discovering knowledge that can have various applications. Reddit.com is an online community and social media platform, which allows Redditors (registered users of Reddit.com) to form groups/subreddit to discuss specific topics. Groups such as \textit{``r/OpioidRecovery''}, \textit{``r/Opiates''} and\textit{ ``r/drugs''} aim to provide people suffering from OUD and seeking remission with psychological support. This function spurs people with OUD to turn to social media for help, posting various problems and confusion. In this way, social media has become a widely utilized form of social support, which is an important recovery resource that can buffer the negative effects of stress on the quality of life of those with OUD~\cite{laudet2008socialsupport}. Indeed, many people with OUD-related problems that have managed to abstain from opioid use are often willing to share their experience with others online. For instance, a Redditor posted on Reddit that ``I was the same way, I only got clean cuz I didn't wanna go back to jail tbh, and had to go cold turkey in rehab''. This study is based on data collected from subreddit groups such as \textit{``r/opioid''}, \textit{``r/opioidrecovery''}, and \textit{``r/drugs''} from Reddit.com in 2018.

\subsection{Dataset}

The dataset used in this paper is collected from Reddit.com, a social news aggregation, web content rating, and discussion website. We developed a web crawling tool using PRAW (Python Reddit API  Wrapper) to collect data. The dataset for this paper is from three subreddits (Subreddits are subsidiary threads or categories within the Reddit website) on Reddit.com, including \textit{``r/opiates"}, \textit{``r/opiatesRecovery"} and \textit{``r/drugs"}. In a subreddit, there are a series of posts that include a post topic and a collection of related comments. 
The dataset consists of two parts. The first consists of 3,000 posts with the top (collected in April 2018) 1,000 posts from each subreddit, since 1,000 is the maximum number that is allowed by PRAW API. This first part includes posts and interactions within posts. After collecting the subreddit data, we extracted a list of Redditor IDs of participants. The second part of the dataset consisted of personal data, which was collected based on the previously mentioned Redditor list that was extracted. 

Data labeling then involved manually reading the posts/comments and categorizing each Redditor into a corresponding group. With intent to implement binary classification, each Redditor was marked with a label and different labels were mutually exclusive. This classification ensured that each participant was assigned to one and only one label. 

More specifically, in the group classification process, we labeled a Redditor ID by manually reading his/her post history in the collected dataset. This labeling process was done by several health science researchers, and labeling agreement among the researchers was required to reduce labeling errors that might otherwise have resulted from different labeling metrics. For the OUD classifier, we followed DSM-5 criteria~\cite{tarrahi2015latent}, in which a Redditor is labeled with OUD if at least two of the listed criteria were met within the past 12 months. For instance, a Redditor was labeled with OUD if he/she admitted that he/she often takes in larger amounts of opioids and/or over a long period of time than intended, and also mentioned a persistent desire or unsuccessful efforts to cut down or control use, or had strong cravings to use. If less than two symptoms were detected, a Redditor was classified to a ``Non-OUD" group. Redditors who met at least two DSM-5 criteria but showed no intentions of limiting or abstaining from use were classified to an ``OUD" group. For instance, an OUD Redditor posted ``Late Christmas Gift for myself 5x 40mg Oxycodone (Toroxycon)''. Redditors were classified into a ``positive recovering OUD'' group if he/she was attempting and/or struggling to seek treatments that reduced symptoms associated with previously met DSM-5 criteria in the last 12 months. For instance, a positive recovering OUD posted ``what to say when returning to work from drug treatment''. If there was no evidence that a Redditor was seeking treatment, this Redditor was classified to a ``non-positive recovering OUD'' group. Similarly, Redditors were classified to an ``OUD relapse'' group if they first classified as in recovery but indicated that they had used opioids again (no matter how many times) in the fifty days previous to the latest post/comment. For instance, a relapsed Redditor posted ``45 Days clean and now I'm drunk and high''. Otherwise, Redditors in recovery were assumed to have stayed clean. 

When classifying participants into the ``positive recovering OUD'' and ``non-positive recovering OUD" groups, one interesting finding was that many participants with OUD had varying degrees of willingness to stay clean, which is consistent with the literature~\cite{bradshaw2013readinessfactors}. That is, ``positive recovering OUD'' group members showed the intention of reducing symptoms associated with DSM-5 criteria while ``non-positive recovering OUD'' group members show less intention of such symptom reduction.

In summary, we collected 1,000 Reddit posts respectively from \textit{``r/opiates''}, \textit{``r/opiatesRecovery''}, and \textit{``r/drugs''}. By utilizing Redditor IDs extracted from the 1,000 posts, we also collected the latest 1,000 comments/posts Reddit API restricts the number of posts that can be collected to 1,000 for each Redditor ID respectively, and obtained a personal post dataset.

\subsection{Redditor Classification}
The Redditor classification is based on the text features. Give a set of Redditors $A=\{a_1, a_2,a_3,\dots\}$, and their corresponding feature variable $F_j=<X_1,\dots,X_m>$, which is represented by these features extracted from text, and their status label  $Y_j \in [0,1]$, the Redditor classification task is to learn a function $f:F_j \to Y_j$, such that the task errors are minimized. Formally, we have
\begin{equation}
    \arg \mathop {\min }\limits_\Psi  \sum\limits_{j = 1}^m {||f({X_1,\dots,X_m}) - {Y_j}{||_2}}
\end{equation}
where $X_j$ and $Y_j$ are a feature variable and label variable respectively, and $\Psi$ is the model parameters learned from training.

Two classifiers were designed and implemented to filter out the research targets: an OUD classifier and a recovering classifier. The first classifier differentiates between OUD and non-OUD, while the second ``within OUD" classifier separates those who are in positive recovery and who hadn't shown any evidence of being in a positive recovery process, and they were denoted as ``positive recovering OUD'' and ``non-positive recovering OUD" group members, respectively. For the OUD classifier, we employed a Support Vector Machine (SVM) classifier that transformed posts/comments into term vector features to determine whether participants had an OUD problem. 
The dataset for this classifier included 1,000 Redditors (419 OUDs and 581 non-OUDs). In the labeled dataset, $70\%$ (namely, $70\%$ of the samples) was used as the training dataset, and the rest were used for testing. 

Once the classifier had been trained, we applied it to the unlabeled dataset. The data of the identified OUDs were then fed into the next classifier to determine whether they were a positive recovering OUD or a non-positive recovering OUD. The second classifier was designed to identify those with OUD who had shown positive attitudes/acts toward seeking recovery. This took another four days to label 1,000 people (with 375 positive recovering OUDs and 625 non-positive recovering OUDs). Among this labeled data, $70\%$ was used as the training dataset and the rest for testing. We show the performance of these two classifiers in Table~\ref{tab:classification}.

\begin{table}[htbp]
  \centering
    \begin{tabular}{lrrrr|rrrr}
    \toprule
    \multicolumn{5}{c|}{OUD Classif.} & \multicolumn{4}{c}{Recorvering Classif.} \\
    \midrule
          & \multicolumn{1}{l}{Acc.} & \multicolumn{1}{l}{Rec. } & \multicolumn{1}{l}{Prec.} & \multicolumn{1}{l|}{F1} & \multicolumn{1}{l}{Acc.} & \multicolumn{1}{l}{Rec. } & \multicolumn{1}{l}{Prec.} & \multicolumn{1}{l}{F1} \\
    LG. & 0.82 & 0.83 & 0.82 & 0.82 & 0.74 & 0.78 & 0.73 & 0.76 \\
    KNN   & 0.70 & 0.81 & 0.76 & 0.7858 & 0.75 & 0.8121 & 0.73 & 0.77 \\
    SVM   & \textbf{0.92} & \textbf{0.94} & \textbf{0.93} & \textbf{0.93} & \textbf{0.88} & \textbf{0.93} & \textbf{0.84} & \textbf{0.88} \\
    \bottomrule
    \end{tabular}%
  \caption{\textbf{Classification results for Logistic Regression, KNN, and SVM.}}
  \label{tab:classification}%
\end{table}%
The SVM classifier is implemented by using the ``Libshorttext'' library. The dataset is processed by converting to lower case, removing punctuation and stop words, tokenizing, stemming and lemmatizing. As shown in Table~\ref{tab:classification}, SVM classifier outperforms logistic regression and KNN in both cases. Moreover, the accuracy of the first classifier was $0.917$, which was better than the accuracy of the second classifier. Thus, there wouldn't be any performance bottleneck if the output of the first classifier was fed into the second classifier.

We manually classified 3,157 people, with 64.9\% of them have OUD problems,  and 35.1\% are non-OUD. However, only 10.4\% of these people with OUDs had shown that they have been positively seeking remission and/or recovery, while the other 89.6\% of those with OUD showed no signs of seeking remission and/or recovery according to our dataset records. Of the recovering Redditors with OUD, 89.3\% will relapse within fifty days.

\begin{figure*}[h]
    \centering
    \includegraphics[width=0.9\linewidth]{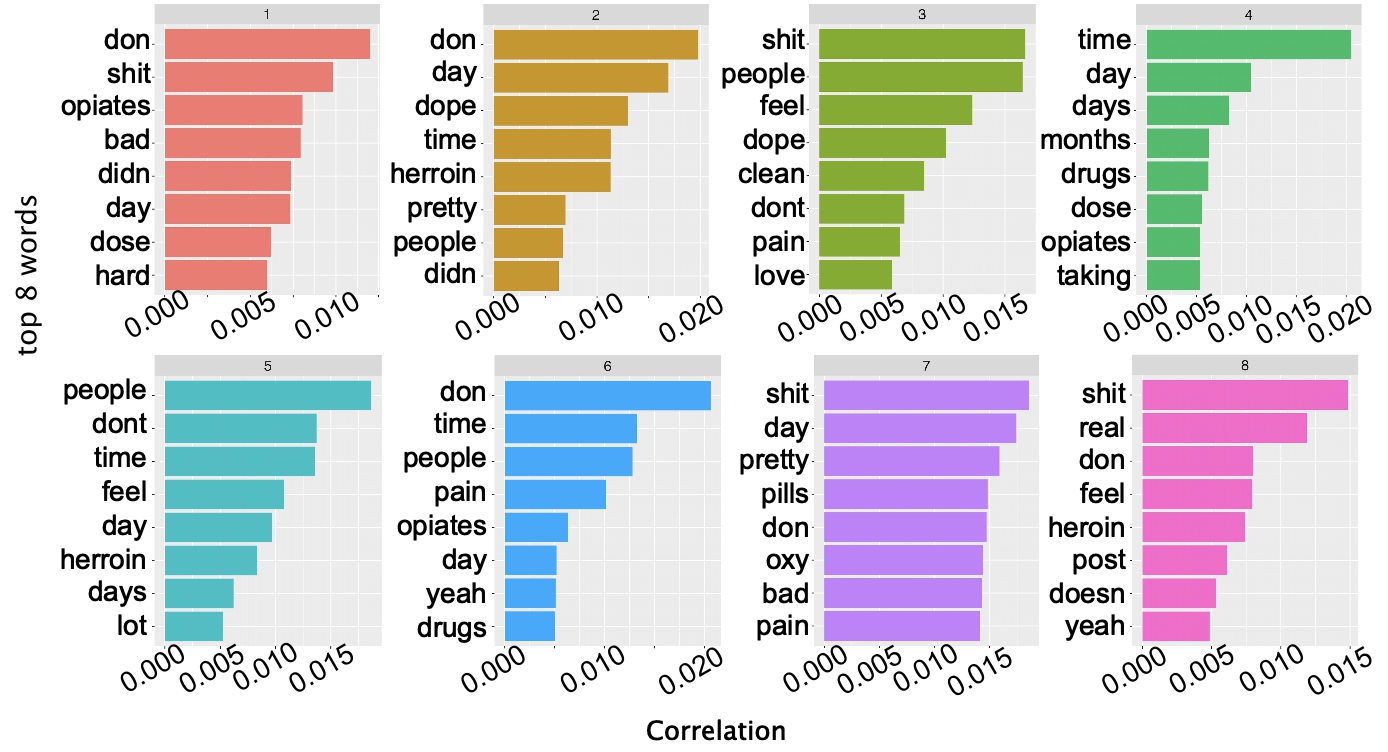}
    \caption{\textbf{Topic modeling results. The topic numbers were set as eight, and top keywords as 8.}}
    \label{fig:topic-map}
\end{figure*}

\subsection{Topic Modeling}

To discover the hidden semantic structures (i.e., ``topics'') that occur in an extensive collection of posts collected from Reddit.com, we first built a topic model to analyze those posts. Given a document in the topic modeling, we would expect particular words to appear in the document more or less frequently. For instance, by specifying two topics, we found the corresponding topics of ``people take opioids because of pain'' and ``people take dope and feels''. Those topics are indicated by words ``people" and ``pain", and ``people'' and ``dope'', respectively, since they appear more frequently in those posts than other words.

According to the Deveaud2014\cite{deveaud2014accurate}, the best topic number is in the range of [4, 11], where model achieve the extremum. Since posts typically concern numerous topics, we specified the existence of multiple topics to further explore any additional potential topics that might emerge. As shown in Figure~\ref{fig:topic-map}, we present the results of an eight-topic model. These topics are ``hard day, don't do opiates'', ``don't dope heroin'', ``people feel love to feel clean'', ``taking drugs or opioids for days'', ``people feel, heroin'', ``people pain, taking opiates or drugs'', ``for pain, don't do oxycodone'', and ``post about heroin'', respectively. In brief, the general topics included 1). Attempts to dissuade people from taking highly addictive opioids (e.g., heroin and oxycodone); 2). Information on how to stay clean; 3). How people feel when they stay clean for an extended period; and 4). Advice about treatment and how to stay clean. 

Even though topic modeling reveals the general topics in certain social communities, individualized user-specific information remains hidden. To expose more details about user-specific patterns for people with OUD that utilize social media, we implemented text co-occurrence and correlation analysis to further dig and discover behavior patterns on social media. 

\subsection{Text Co-occurrence and Correlation}
\begin{figure*}[t]
	\centering
	\subfigure[]{
	    \includegraphics[width= 0.99\textwidth]{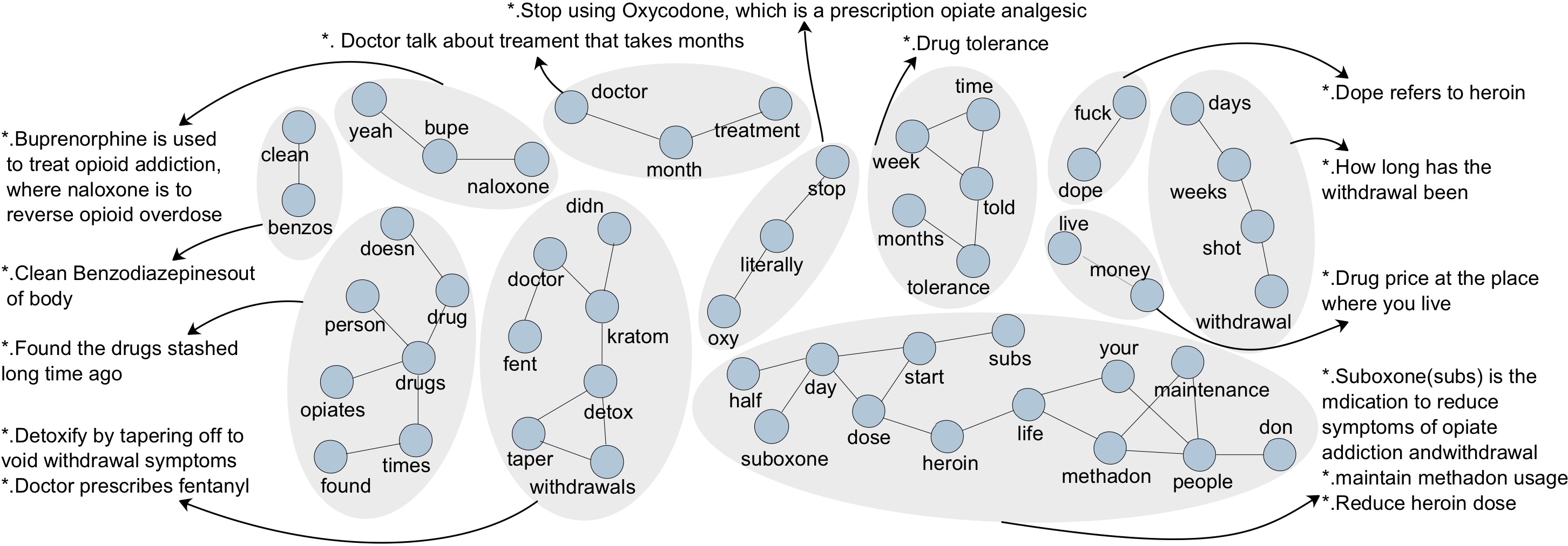}
	}
    \subfigure[]{
		\includegraphics[width= 0.3\textwidth]{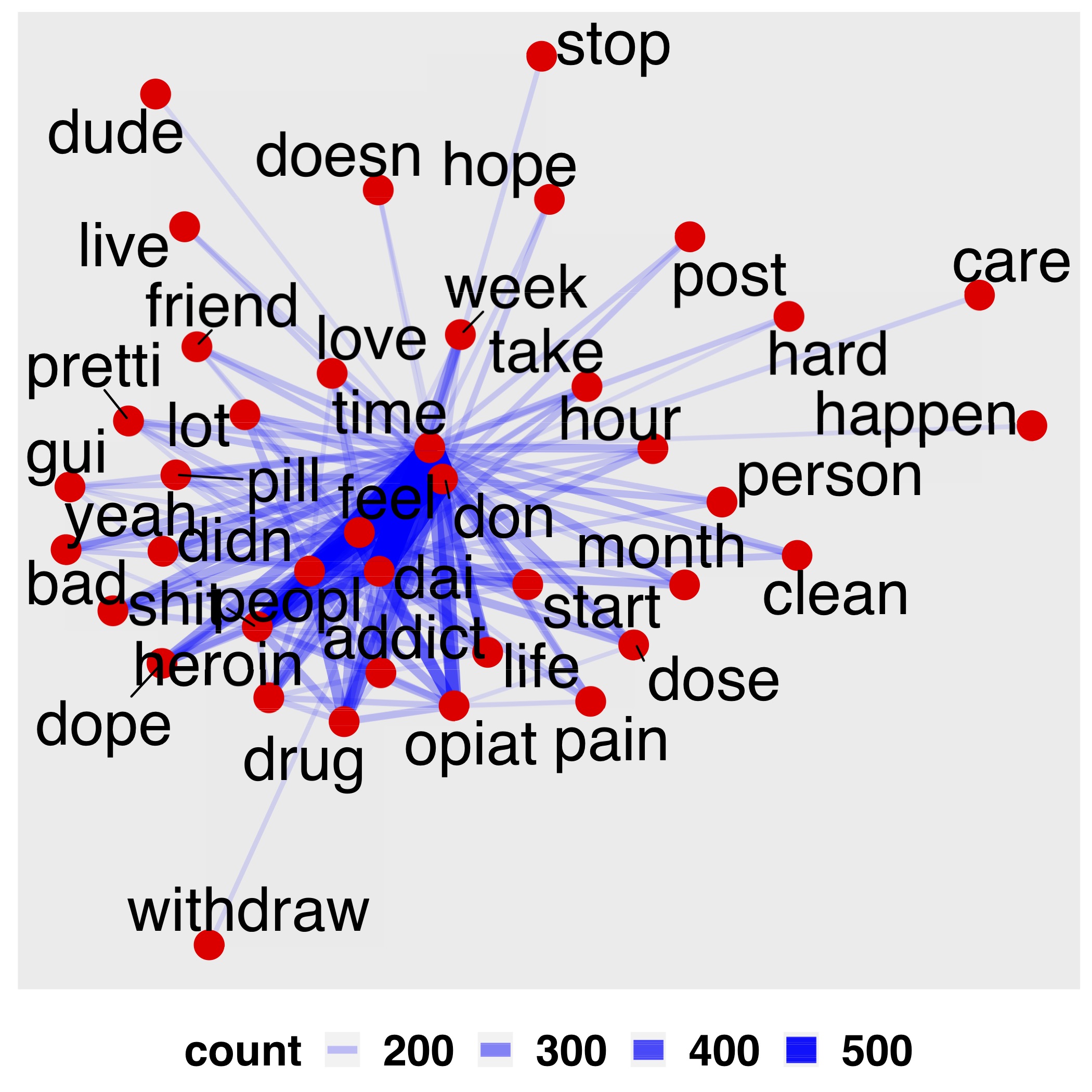}
    }
        \subfigure[]{
		\includegraphics[width= 0.3\textwidth]{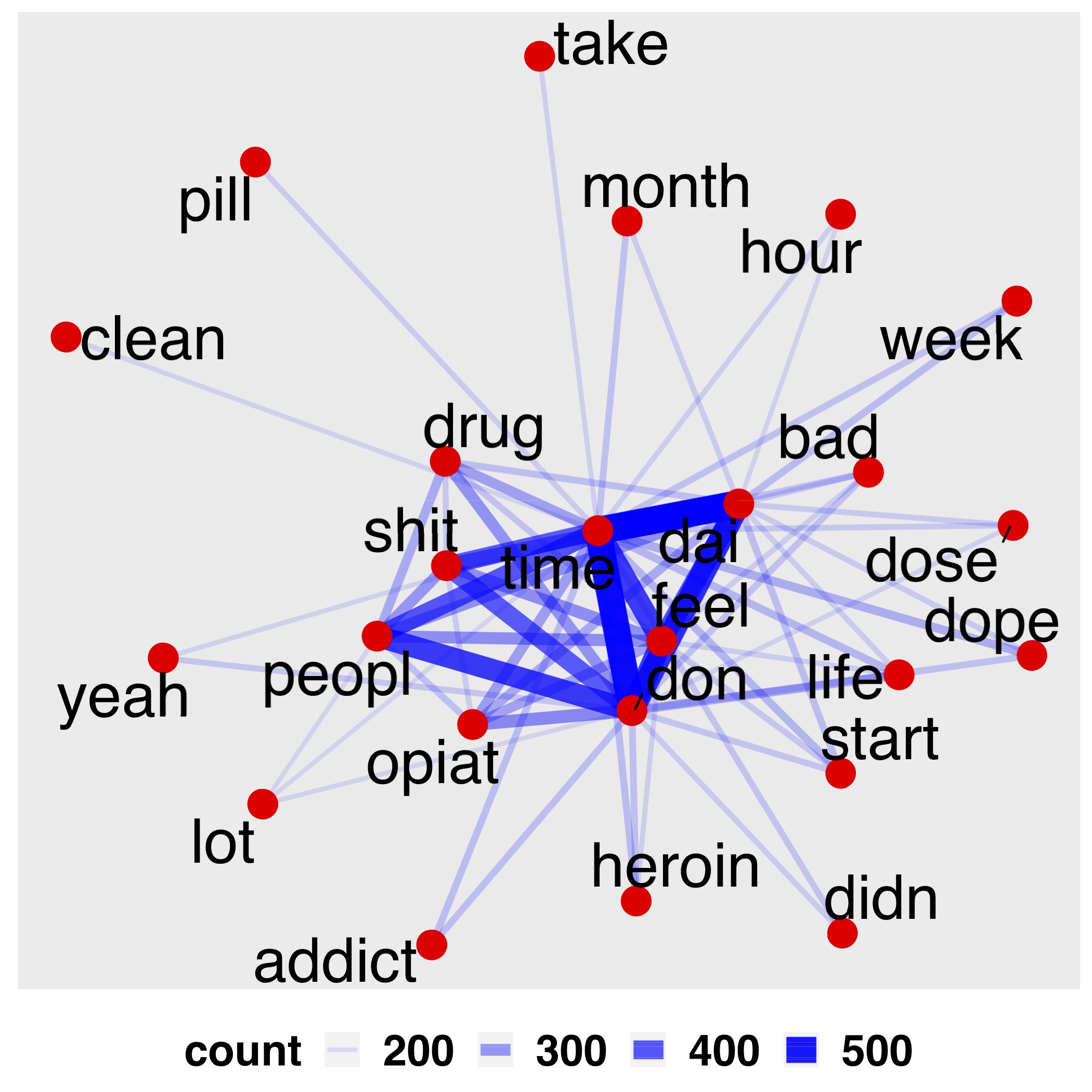}
    }
        \subfigure[]{
		\includegraphics[width= 0.3\textwidth]{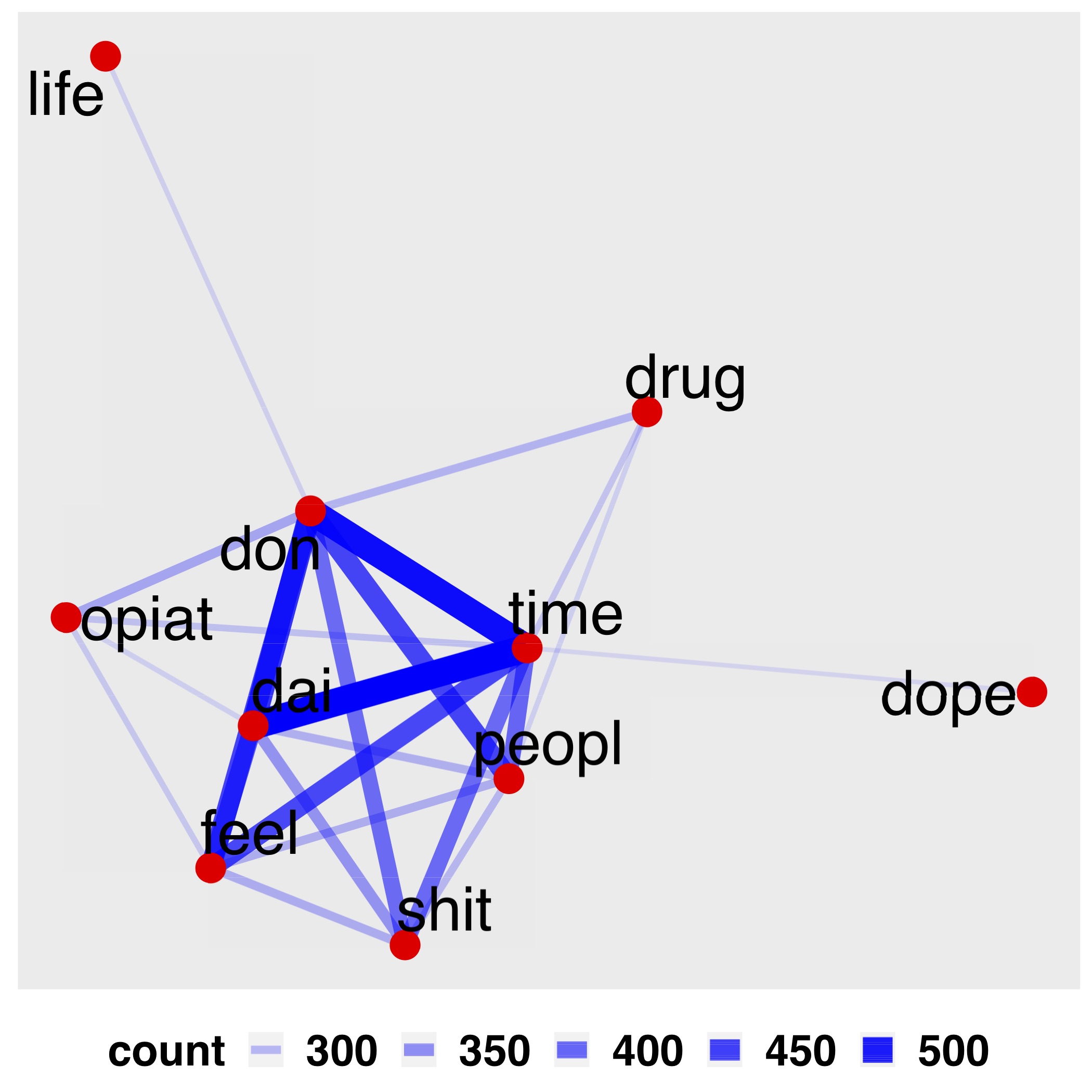}
    }
\caption{\textbf{(a): Word correlation with Phi coefficiency$\geq$0.2. (b): Word bigrams with co-currence co-occurrence count $\geq$ 150. (c): Word bigrams with co-occurrence count $\geq$ 200. (d): Word bigrams with co-occurrence count $\geq$ 250.}}
\label{fig:word_corr_cluster}
\end{figure*}

Figure~\ref{fig:word_corr_cluster} A. shows the word correlation with annotations. The word correlation and bigram illustrate the social interactions between different Redditors. We generated word correlation graphs to obtain a full picture of their conversations. A textual analysis was implemented to capture the relationship between words, to examine words that tend to follow each other immediately, or tend to co-occur within the same posts. Sentences were tokenized into sequences of words. By checking how often a word $X$ was followed by word $Y$, the relationships between words are measured by the Phi efficient metric, and the Phi efficient is defined as
\begin{equation}
    \phi  = \frac{{{n_{11}}{n_{00}} - {n_{10}}{n_{01}}}}{{\sqrt {{n_{1.}}{n_{0.}}{n_{.0}}{n_{.1}}} }},
    \label{equation:phi}
\end{equation}
where $n_{11}$ represents the number of posts where both $X$ and $Y$ appear, $n_{00}$ is the number where neither appears, and $n_{10}$ is the cases where one appears without the other.
In this case, a pair of words with higher phi coefficiency indicates that this word pair appears more frequently than other word pairs do.

By gradually increasing the Phi coefficient, the words that are most likely to appear with each other were detected, as shown in Figure~\ref{fig:word_corr_cluster} (b), (c), and (d). After analyzing a large number of posts from Reddit, the relationship was visualized in Figure~\ref{fig:word_corr_cluster}. By analyzing and annotating these keywords, some frequent topics are summarized as follows:
\begin{itemize}
    \item Talking about detoxification processes and assisting medications such as `naloxone' and `bupe', in conjunction with `doctor' and `detox'. Participants of these discussion topics are Redditors who are motivated and willing to find recovery from their addiction and manage to stay clean. Also, some of these Redditors who have been clean for a longer period participate in those discussions and share their experiences about how to stay clean.
    \item Describing pains and prescriptions from doctors, as indicated by keywords such as `pain', `doctor', and `painkiller'. These topics are consistent with the fact that opioids are often originally used as a prescribed painkiller. Some Redditors do have bodily and/or chronic pain problems and concerns. Though opioids are an effective painkiller and have curative effects, after months or years of taking opioids, these Redditors often develop an opioid dependency.
    \item Sharing of withdrawal symptoms, which symptoms fall into two categories. One category involves mental/psychological symptoms such as anxiety, insomnia, and social isolation. The other category involves physical symptoms, including headaches and dizziness. Both categories of withdrawal symptoms are very uncomfortable and may lead people with OUD to a compulsive relapse if they do not have proper guidance and support.
    \item Talking about what kind of opioids they take, how it feels, and the dosage of drugs they take. These topics are indicated by words such as ``hight'', ``enjoy'', ``mg (the dosage of opioids)'', etc. For those topics, Redditors always share their experiences about what kinds of opioids they are taking, or they have been taking, or they used to take. But they try to be cautious about the dosage because of the risks of death caused by overdose.
\end{itemize}

\begin{figure*}[hbpt]
    \centering
    
    \subfigure[]{
        \includegraphics[width=0.48\linewidth]{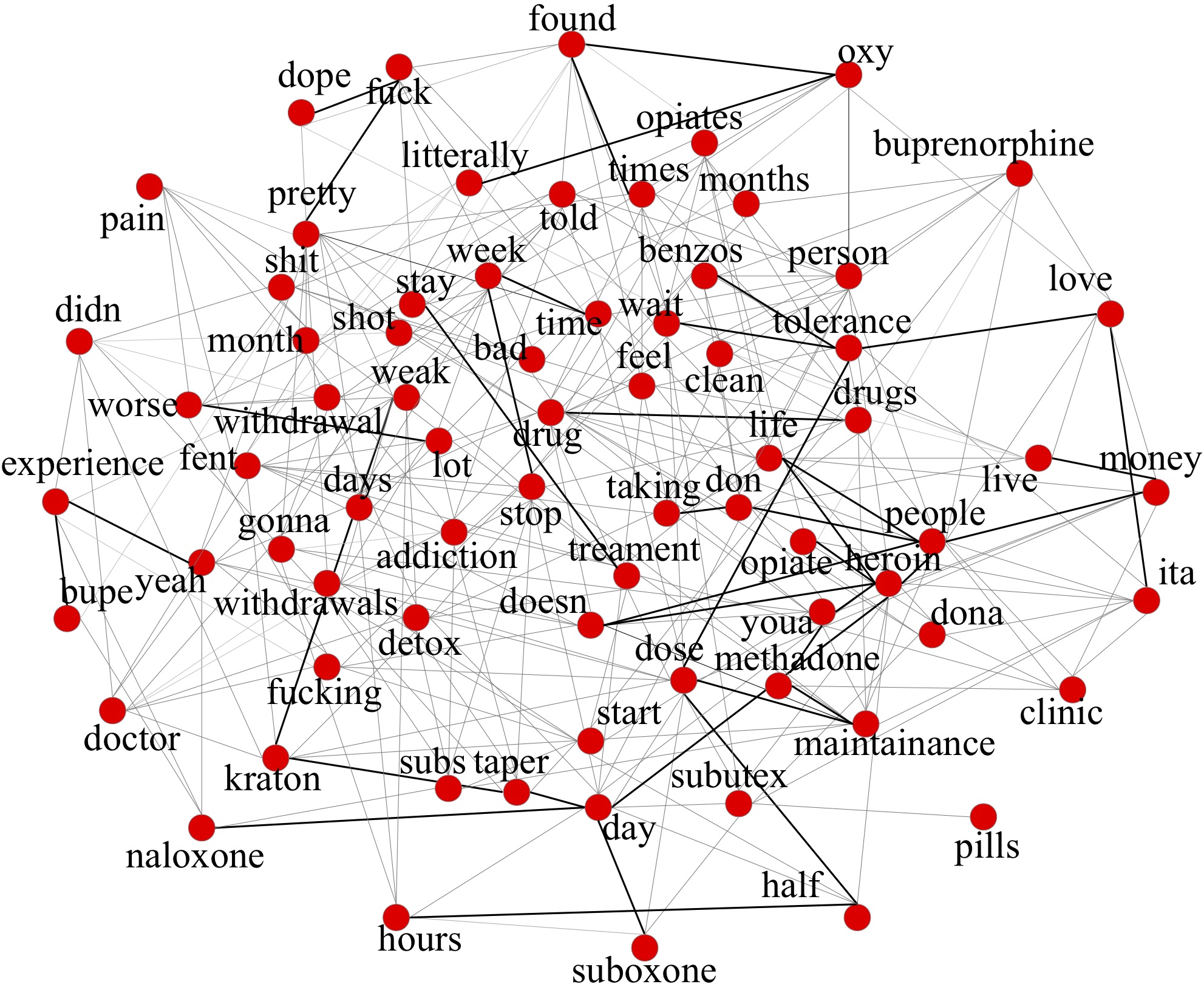}
    }
    \subfigure[]{
        \includegraphics[width=0.48\linewidth]{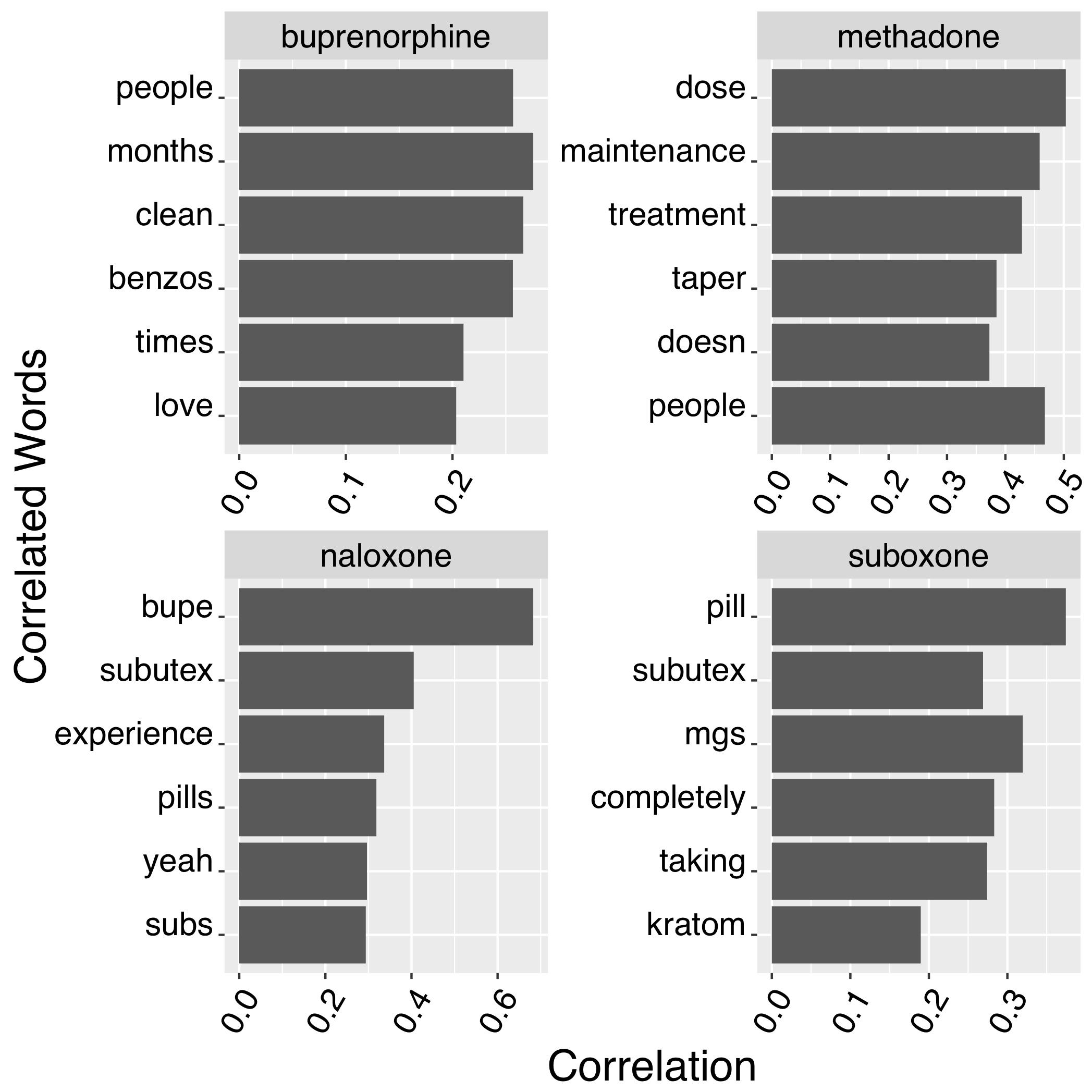}
    }
    
    \caption{\textbf{A: Word correlation (in the same posts) with correlation coefficiency $\geq$ 0.17. B: Top ords that are used when talking about opioid related medicines.}}
    \label{fig:medical}
\end{figure*}

Also, since treatment is critical for people with OUD, figuring out how those with OUD evaluate treatment-assisting medications is very important. By specifying treatment-assisting medications such as `buprenorphine', `methadone', `naloxone', and `suboxone', we detected and filtered out words that were associated with those medicines in these communities. In Figure~\ref{fig:medical} A, we visualize the word correlation on posts that are about treatments. Figure~\ref{fig:medical} B shows the top six words with the highest correlation coefficiency for each medication. As shown in this figure, most of the words related to these medications are neutral, representing neutral feedback from those with OUD. For instance, Buprenorphine is associate with `` people months clean benzos love''. Also, a lot of words about dosage are used such as `dose' and `mgs'. Those words indicate that when they use medications, they seem to be paying much attention to dosage, perhaps because high dosages may have lethal rather than curative effects.

\subsection{Emotion of People with OUD}

Relapses for persons with OUD are associated with younger age, heavy use before treatment, a history of injecting, and not following up with aftercare~\cite{smyth2010lapse}. They are also frequently associated with extreme emotions and other stressors~\cite{elster2009strong}.
Specifically, people seeking remission or rehabilitation can experience extreme emotions that range from great highs to great lows, and sometimes they feel like their emotions are out of control. This is important as the well-accepted Self-Medication Hypothesis~\cite{Khantzian1985smh} proposes that people often use substances to self-medicate the negative and painful emotions associated with trauma, stress, and/or comorbid mental health disorders~\cite{Tronnier2015regulationsmh}. To explore emotions that are associated with relapse, an emotion analysis was implemented to show the observational results.

On Reddit.com, people with OUD are free to express their joy and sadness, trust and disgust, and anticipation and anger, etc. in their posts or comments. These sentiments, along with their other online behaviors, could disclose information about their remission/recovery status and thus serve as relapse risk indicators and therefore also for the need for intervention that might assist with relapse prevention~\cite{yang2019sstd}. We therefore sought to capture subtle emotional sentiments from those with OUD and associate these sentiments with OUD associated behaviors such as staying clean or relapsing. By studying Redditor posts as well as their interactions with other Redditors, it was hoped that identified emotions can increase understanding of remission/recovery status. Such understanding may ultimately lead to transforming the sentiments and behavior patterns that emerge from Redditor posts into possible indicators of relapse risk and the need for some type of intervention to prevent relapse.

A word-emotion association lexicon~\cite{mohammad2013crowdsourcing} was utilized to study the emotional content of each comment from a Redditor. In the sentiment analysis, we focused on dominant rather than submerged emotions. For instance, a post may include a complex emotion that is a combination of several emotions. In this case, a comment is labeled with the emotion that has the highest score, which is defined in equation~\ref{emotionCount}, ~\ref{emotionCountNorm} and ~\ref{emotionWord}.

In particular, suppose $L_i$ is the subset of a lexicon which contains words in emotion $i$. The emotion count $emotion\_count\_i$ and normalized count $n\_emotion\_count\_i$ for the emotion $i$ are: 

\begin{equation}
    \label{emotionCount}
    emotion\_count\_i = \sum\limits_j{word_{ij}}
\end{equation}
\begin{equation}
    \label{emotionCountNorm}
    n\_emotion\_count\_i = \frac{emotion\_count\_i}{max(emotion\_counts)}  
\end{equation}

Where:
\begin{equation}
    \label{emotionWord}
  word_j =
  \begin{cases}
    1 & \text{if $word_j$ in $L_i$} \\
    0 & \text{otherwise}
  \end{cases}
\end{equation}

\begin{figure*}[htp]
    \centering
    \includegraphics[width=0.9\linewidth]{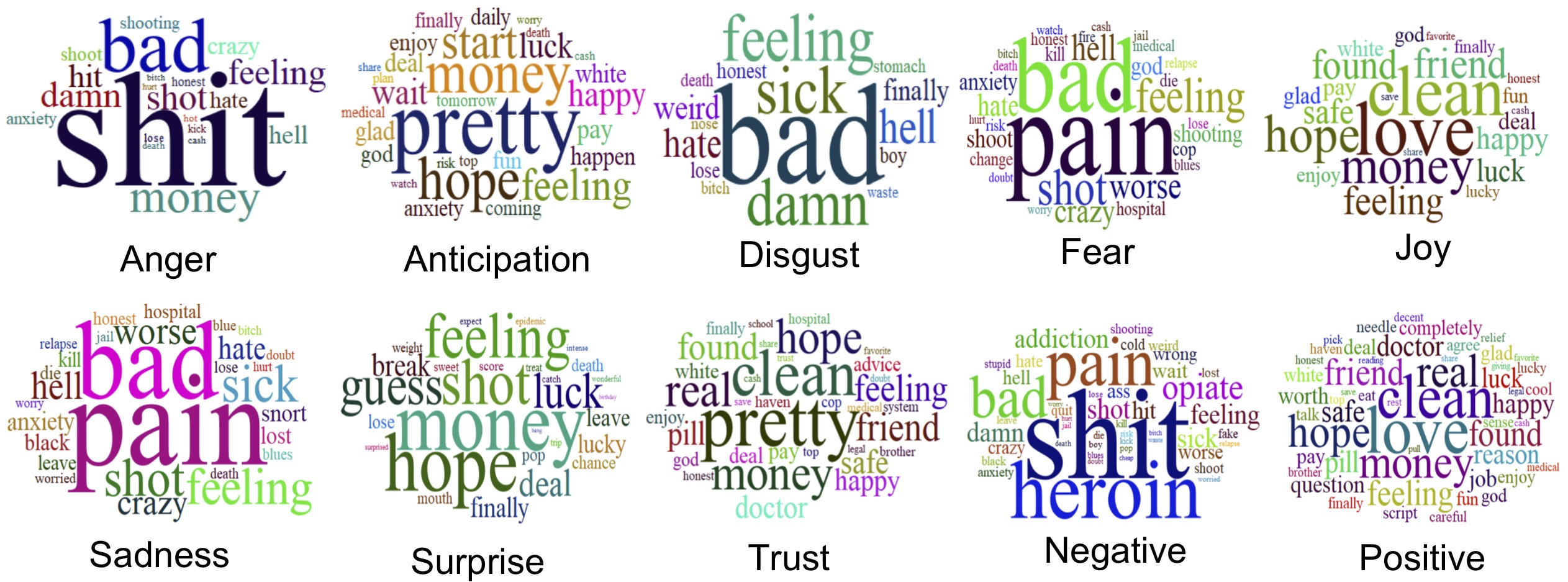}
    \caption{\textbf{Word cloud with ten different emotions. The drug names such as Heroin, OxyContin and Vicodin are not the dominant words in the wordcloud, since people talk more about their feelings instead of the drug itself. Also, stopwords are removed from the wordcloud.}}
    \label{fig:wordcloud}
\end{figure*}

Ten categories of emotion were identified as shown in Figure~\ref{fig:wordcloud}: anger, anticipation, disgust, fear, joy, sadness, surprise, trust, negative, positive. Also consistent with addiction literature~\cite{randles2013shamerelapse}, we found that people with OUD can be highly influenced by their emotions. Words that express emotions or feelings, such as ``feeling'', ``bad'', and  ``high'', are repeatedly used. Among the 670 persons with OUD who had relapsed, 72\% showed the predominant emotions that were `negative' or that of `joy'. This observation shows relapse is highly related to more extreme emotions such as `negative' and `joy' based-emotions.

\section{Discussion}
The opioid crisis is one of the leading threats to public health in the U.S., and therefore treatment admissions are rising. Despite positive benefits from treatment, many persons seeking remission relapse, and such relapse is difficult to predict. Nevertheless, a thorough understanding of social media behavior patterns that associate with remission can promote potential supportive and even potential relapse prevention-based interventions. In this paper, we conducted a series of experiments and observed significant Opioid addiction patterns by analyzing posts and comments collected from Reddit.com. These advanced analyses demonstrate 1) how information from social media posts/discussions can be used to more accurately predict potential relapse, 2) how discussions might be discovered around various topics that association with use, treatment, and remission, and 3) how emotional/affective sentiments from these discussions may be used to predict relapse risk.

First, only 10.4\% of those identified with OUD actively seek remission/recovery, while the 89.6\% show no indication of seeking recovery. This demonstrates the important need that the majority of those with OUD to progress through various motivational stages that may yield a greater readiness for change. Additionally, the machine-learning based classifiers used in this study showed high accuracy and precision in predicting group classification outcomes.
Second, people with OUD who are seeking remission/recovery more generally have positive judgments regarding medications such as Buprenorphine, Methadone, Naloxone, and Suboxone; and they pay close attention to dosage. Thus, this finding supports that medication assisted treatment (MAT; use of medications such as Buprenorphine and Methadone) may be helpful to those seeking recovery from OUD. This is important as MAT is often a controversial topic and widely underutilized resource in substance use disorder treatment~\cite{Knudsen2011MATImplementation}. Third, 72\% of those seeking recovery from OUD - but appear to have relapsed - showed both dominant `negative' as well as `joy' based emotions that associated with relapse. Therefore, we infer people with OUD may experience a wide range of emotion and emotional shifts, and that without support that aids emotional regulation they have higher relapse risk.
In sum, these multi-disciplinary analyses conducted in this paper help disclose several behavioral patterns and characteristic features of persons with OUD, making it more possible to detect and identify precursors of relapse, increasing assessment and prediction of relapse risk. Ultimately, this may lead to the development and implementation of personalized OUD-based interventions that enhance relapse prevention and OUD remission.

\bibliographystyle{IEEEtran}

\bibliography{reference}




\end{document}


\begin{table*}[htbp]
  \centering
    \begin{tabular}{|p{0.25em}ll|}
    \toprule
    \multicolumn{3}{|p{60.92em}|}{An opioid use disorder is defined as a problematic pattern of opioid use that leads to serious impairment or distress. Doctors use a specific set of criteria to determine if a person has a substance use problem. To be diagnosed with an opioid use disorder, a person must have 2 or more of the following symptoms within a 12-month period of time. } \\
    \midrule
    \multicolumn{3}{|p{60.92em}|}{\textbf{Loss of Control}} \\
    \midrule
    \multicolumn{1}{|l|}{1} & \multicolumn{1}{p{23em}|}{{Substance taken in larger amounts or for a longer time than intended}} & \multicolumn{1}{p{28em}|}{{I did not mean to start using so much.}} \\
    \midrule
    \multicolumn{1}{|l|}{\textcolor[rgb]{ .18,  .176,  .196}{\textbf{2}}} & \multicolumn{1}{p{23em}|}{\textcolor[rgb]{ .18,  .176,  .196}{Persistent desire or unsuccessful effort to cut down or control use of a substance}} & \multicolumn{1}{p{28em}|}{\textcolor[rgb]{ .18,  .176,  .196}{I have tried to stop a few times before, but I start using this drug again every time.}} \\
    \midrule
    \multicolumn{1}{|l|}{\textcolor[rgb]{ .18,  .176,  .196}{\textbf{3}}} & \multicolumn{1}{p{23em}|}{\textcolor[rgb]{ .18,  .176,  .196}{Great deal of time spent obtaining, using, or recovering from substance use}} & \multicolumn{1}{p{28em}|}{\textcolor[rgb]{ .18,  .176,  .196}{Everything I do revolves around using this drug. (In severe cases, most/all of a person's daily activities may revolve around substance use.)}} \\
    \midrule
    \multicolumn{1}{|l|}{\textcolor[rgb]{ .18,  .176,  .196}{\textbf{4}}} & \multicolumn{1}{p{23em}|}{\textcolor[rgb]{ .18,  .176,  .196}{Craving (a strong desire or urge) to use opioids}} & \multicolumn{1}{p{28em}|}{\textcolor[rgb]{ .18,  .176,  .196}{I wanted to use so badly, I couldn't think of anything thing else.}} \\
    \midrule
    \multicolumn{3}{|l|}{\textbf{Social Problems}} \\
    \midrule
    \multicolumn{1}{|l|}{\textcolor[rgb]{ .18,  .176,  .196}{\textbf{5}}} & \multicolumn{1}{p{23em}|}{\textcolor[rgb]{ .18,  .176,  .196}{Continued opioid use that causes failures to fulfill major obligations at work, school, or home}} & \multicolumn{1}{p{28em}|}{\textcolor[rgb]{ .18,  .176,  .196}{I keep having trouble at work/have lost the trust of friends and family because of using this drug.}} \\
    \midrule
    \multicolumn{1}{|l|}{\textcolor[rgb]{ .18,  .176,  .196}{\textbf{6}}} & \multicolumn{1}{p{23em}|}{\textcolor[rgb]{ .18,  .176,  .196}{Continued opioid use despite causing recurrent social or personal problems}} & \multicolumn{1}{p{28em}|}{\textcolor[rgb]{ .18,  .176,  .196}{I can not stop using, even though it's causing problems with my friends/family/boss/landlord.}} \\
    \midrule
    \multicolumn{1}{|l|}{\textcolor[rgb]{ .18,  .176,  .196}{\textbf{7}}} & \multicolumn{1}{p{23em}|}{\textcolor[rgb]{ .18,  .176,  .196}{Important social, occupational, or recreational activities are reduced because of opioid use}} & \multicolumn{1}{p{28em}|}{\textcolor[rgb]{ .18,  .176,  .196}{I have stopped seeing my friends and family, and have given up my favorite hobby because of drugs.}} \\
    \midrule
    \multicolumn{1}{|l|}{\textbf{Risky Use}} & \multicolumn{1}{l|}{} &  \\
    \midrule
    \multicolumn{1}{|l|}{\textcolor[rgb]{ .18,  .176,  .196}{\textbf{8}}} & \multicolumn{1}{p{23em}|}{\textcolor[rgb]{ .18,  .176,  .196}{Recurrent opioid use in dangerous situations}} & \multicolumn{1}{p{28em}|}{\textcolor[rgb]{ .18,  .176,  .196}{I keep doing things that I know are risky and dangerous to buy or use this drug.}} \\
    \midrule
    \multicolumn{1}{|l|}{\textcolor[rgb]{ .18,  .176,  .196}{\textbf{9}}} & \multicolumn{1}{p{23em}|}{\textcolor[rgb]{ .18,  .176,  .196}{Continued opioid use despite related physical or psychological problems}} & \multicolumn{1}{p{28em}|}{\textcolor[rgb]{ .18,  .176,  .196}{I know that using this drug causes me to feel badly/messes with my mind, but I still use anyway.}} \\
    \midrule
    \multicolumn{3}{|p{35.92em}|}{\textbf{Pharmacological Problems}} \\
    \midrule
    \multicolumn{1}{|l|}{\textcolor[rgb]{ .18,  .176,  .196}{\textbf{10}}} & \multicolumn{1}{p{23em}|}{\textcolor[rgb]{ .18,  .176,  .196}{Tolerance (the need to take higher doses of a drug to feel the same effects, or a reduced effect from the same amount)}} & \multicolumn{1}{p{28em}|}{\textcolor[rgb]{ .18,  .176,  .196}{I have to take more and more of the drug to feel the same high.}} \\
    \midrule
    \multicolumn{1}{|l|}{\textcolor[rgb]{ .18,  .176,  .196}{\textbf{11}}} & \multicolumn{1}{p{23em}|}{\textcolor[rgb]{ .18,  .176,  .196}{Withdrawal (the experience of pain or other uncomfortable symptoms in the absence of a drug)}} & \multicolumn{1}{p{28em}|}{\textcolor[rgb]{ .18,  .176,  .196}{When I stop using the drug for a while, I am in a lot of pain.}} \\
    \midrule
    \multicolumn{3}{|p{35.92em}|}{\textcolor[rgb]{ .18,  .176,  .196}{\textbf{Summary:  OUD: $N_{symptoms} \ge 2$, \; Non-OUD:$N_{symptoms} < 2$ }}} \\
    \bottomrule
    \end{tabular}%
    \caption{Labeling criteria and text examples from DSM-5. A Redditor is labeled with OUD if at least two symptoms are observed. It's used for labeling OUD and non-OUD, and positive recovering OUD and non-positive recovering OUD.}
  \label{tab:addlabel}%
\end{table*}%